\documentclass[12pt,preprint]{aastex}
\usepackage{graphicx}
\usepackage{ulem}
\usepackage{rotating}
\usepackage{lscape}

\title{Anachronistic Grain Growth and Global Structure of the Protoplanetary Disk Associated with the Mature Classical T Tauri Star, PDS 66}

\author{Stephanie R. Cortes$^{1}$, Michael R. Meyer$^{1}$, John M. Carpenter$^{2}$, Ilaria Pascucci$^{1,5}$, Glenn Schneider$^{1}$, Tony Wong$^{3}$, Dean C. Hines$^{4}$}
\affil{$^{1}$Steward Observatory, The University of Arizona, 933 N Cherry Ave., Tucson, AZ, 85721; scortes@as.arizona.edu, mmeyer@as.arizona.edu, pascucci@as.arizona.edu, gschneider@as.arizona.edu\\
$^{2}$Department of Astronomy, California Institute of Technology, Pasadena, CA 91125; jmc@astro.caltech.edu\\
$^{3}$Astronomy Department, University of Illinois, 1002 W. Green Street, Urbana, IL 61801; wongt@uiuc.edu\\
$^{4}$Space Science Institute, 4750 Walnut Street, Suite 205, Boulder, CO 80301; dhines@as.arizona.edu\\
$^{5}$Department of Physics \& Astronomy, Johns Hopkins University, 366 Bloomberg Center, 3400 N. Charles Street, Baltimore, MD 21218; pascucci@pha.jhu.edu}

\begin{document}

\begin{abstract}
We present ATCA interferometric observations of the old (13 Myr), nearby (86pc) classical T Tauri star, PDS 66.  Unresolved 3 and 12 mm continuum emission is detected towards PDS 66, and upper limits are derived for the 3 and 6 cm flux densities.  The mm-wave data show a spectral slope flatter than that expected for ISM-sized dust particles, which is evidence of grain growth.  We also present HST/NICMOS 1.1 micron PSF-subtracted coronagraphic imaging observations of the circumstellar environment of PDS 66. The HST observations reveal a bilaterally symmetric circumstellar region of dust scattering $\sim$0.32\% of the central starlight, declining in surface brightness as r$^{-4.53}$.  The light-scattering disk of material is inclined 32$\pm$5$^{\circ}$ from a face-on viewing geometry, and extend to a radius of 170 AU.  These data are combined with published optical and longer wavelength observations to make qualitative comparisons between the  median Taurus and PDS 66 spectral energy distributions (SEDs).  By comparing the near-infrared emission to a simple model, we determine that the location of the inner disk radius is consistent with the dust sublimation radius ($\sim$1400 K at 0.1 AU).  We place constraints on the total disk mass using a flat-disk model and find that it is probably too low to form gas giant planets according to current models.  Despite the fact that PDS 66 is much older than a typical classical T Tauri star ($\le$5 Myr), its physical properties are not much different.
\end{abstract}

\section{Introduction}\label{intro}
Classical T Tauri stars (cTTS) are low mass, pre-main sequence stars that are surrounded by accretion disks, believed to be an integral part of star formation process.  Much of the cloud material from which the star forms is thought to pass through the disk and accrete onto the star \citep{shu87}.  PDS 66, discussed in detail in this paper, was identified as a cTTS by \cite{GH}.  The subsequent stage in the pre-main sequence evolution scenario \citep[e.g.][]{lada87} is weak-emission T Tauri, when signs of active accretion are no longer observed.  Infrared excess studies in young clusters indicate that disks dissipate on $\sim$10 Myr timescales \citep{haisch01}, although there are some examples of disks that still contain their primordial gas at or beyond 10 Myr (e.g. TW Hya, CQ Tau, St 34, Hen 3-600).

Several processes affect protoplanetary disk evolution, important among them:  (1) initial conditions in the cloud, such as the reservoir of material, the initial angular momentum, and the magnetic field; (2) the conservation of angular momentum which leads to disk spreading as accretion takes place \citep{hartmann98}, therefore the size of the disk likely depends on its age; (3) grain growth and fragmentation affects the dust opacity, and therefore the observed SED; (4) the presence of a planet is expected to create gaps in the disk, leading to minima in the SED which cannot be accounted for by other means; (5) disk dissipation via stellar winds or photoevaporation \citep[e.g][]{holl05} will dictate how long a disk is observable.  It is of current interest to study disks over a range of ages to test current evolutionary models.

The combination of infrared and sub-mm observations provide a wealth of information about protoplanetary disks.  The near-infrared emission above stellar photospheric levels, or lack thereof, provides information about the inner edge of the disk \citep[e.g.][]{calvet}.  The spectral shape of the silicate feature at 10 $\micron$ is a probe of grain growth and crystalline fraction of silicates that must have formed within the disk \citep[e.g.][]{kemper, bouw08}.  The compositional differentiation of the dust population within the disk is a relic of the temperature history during epochs of cloud collapse and early disk evolution.  The slope of the far-infrared emission is affected by the amount of disk flaring (radial dependence of disk scale height above the midplane).  If the disk is geometrically flatter, then the longer-to-shorter wavelength flux density ratio \citep[e.g. 70$\micron$/30$\micron$, ][]{bouw08} will be relatively lower than if the disk is more flared.  Finally, the slope of the millimeter emission is sensitive to grain growth, in that as grains become larger, they will contribute more opacity at longer wavelengths.  Therefore, an evolving grain size distribution will result in a different slope in the millimeter spectrum as grains grow to millimeter sizes and larger.  The earliest stages of planet formation require growing micron-sized grains into centimeter-sized pebbles, and on up to meter-sized rocks.  Observational evidence that this process takes place is becoming more common, and the presence of cm-sized grains has been inferred in a few circumstellar disks \citep[e.g.][]{wilner05}.  

PDS 66 (MP Mus, Hen 3-892), a K1IVe cTTS, is a member of Lower Centaurus Crux (LCC), a subgroup of the Sco Cen OB association \citep{MML02}, exhibiting excess infrared emission from 3 to 70 $\micron$ that has been interpreted as arising from a primordial circumstellar disk \citep[e.g.,][]{GH, silv06}.  \cite{hill08} presented the 3 to 70 $\micron$ SED, fit to single-temperature photosphere plus disk blackbody curves (their Figure 6), and interpreted the emission as arising from a wide range of temperatures.  The age of the LCC association has been determined by placing the known members of LCC on an HR diagram and using theoretical evolutionary tracks.  \cite{MML02} estimated the distance to PDS 66 (86 pc) and the age for LCC (17$\pm$5 Myr) based on three evolutionary models. Recent work by \cite{PM}, which includes more members of LCC, suggests a spatial dependence of the ages, such that the northern region appears to be older and the southern part younger.  PDS 66 is in the southern part of LCC which appears to have an age of $\sim$13 Myr.  The periodic photometric variation of PDS 66 at visible wavelengths is a signature of stellar rotation, since it is thought that accretion onto the star produces observable hot spots that co-rotate with the star \citep{bouv93, edwards, kund06}.  \cite{bat} find that PDS 66 has a rotation period of 5.75  $\pm$ 0.03 days.  PDS 66 is still actively accreting at an age that is substantially older than the median age for cTTS.  \cite{pasc07} derived a mass accretion rate for PDS 66 of 5$\times$10$^{-9} M_{\sun}$/yr by modeling the H$\alpha$ line profile; the presence of accreting gas confirms that this is still a primordial disk.

Here we report new multi-wavelength observations combined with information extant in the literature to better characterize the disk geometry and properties of the circumstellar material in the PDS 66 proto-planetary disk system.  This work offers constraints against which future models of the system can be tested, and that will eventually reveal the similarities and differences between PDS 66 and other cTTS.  In \S \ref{obs}, we describe the observations and calibration of our Australia Compact Telescope Array data, the near-IR imaging observations using the Near Infrared Camera and Multi-Object Spectrometer (NICMOS) coronagraph on the Hubble Space Telescope (HST), followed by a presentation of the SED composed mostly of data that we extract from the literature.  \S \ref{scatanalysis} describes the disk parameters that we derive from the near-IR scattered light observations.  \S \ref{results} describes the constraints on the inner disk properties, derived by fitting models to the NIR region.  We discuss our results for PDS 66 in the context of other T Tauri stars in \S \ref{discuss}, as well as the potential for planet formation within the disk.  We finish with a summary of our results in \S \ref{summ}.

\section{Observation and Data Reduction}\label{obs}

\subsection{NICMOS Data}
Hubble Space Telescope (HST) imaging observations of PDS 66 were obtained as part of a Near Infrared Camera and Multi-Object Spectrometer (NICMOS) coronagraphic imaging survey (HST/GO 10527; PI: Hines) of a Spitzer-selected sample of 41 stars exhibiting strong infrared excesses.  The observational strategy for all target stars in the GO 10527 sample is identical and is described in detail by \cite{hines07}. 

PDS 66 was observed at two celestial field orientations (spacecraft roll angles) to enable differentiating between artifacts in the point spread function (PSF) from true circumstellar features \citep[e.g.][]{schn01}.  The PDS 66 coronagraphic observations (r = 0.3$^{\prime\prime}$ image plane mask with NICMOS camera 2; image scale = 75.8 mas/pixel) obtained on 13 April 2006\footnote{Visit numbers 19 and 20 as per the Multimission Archive at Space Telescope http://archive.stsci.edu/hst/search.php} utilizing two contiguous HST orbits. In each orbit the target was autonomously positioned into the coronagraph following short (0.915 s) unocculted target acquisition images in the F171M filter. These images were also used in conjunction with the spacecraft positioning information to determine the position occulted ($<$ 5 mas RSX differential uncertainty).  Each orbit resulted in 2591 s of total coronagraphic integration time with the F110W filter. At the end of each orbit, PDS 66 was directly imaged (non-coronagraphically) with four dither offset 6 s exposures to enable flux density normalization for PSF subtraction. 

Images derived of PDS 66 and template PSFs (discussed below) were created following the method of \cite{schn05} (SSH05) and references therein. Within each visit, the individual images were median combined after verifying their repeatabilities. A suite of PSF subtracted PDS 66 images were created from the target and candidate PSF template median combined images from an ensemble of bright GO/10527 spectral type G-F targets.  Initial target:template flux scaling was performed by using 2MASS catalog J (a close surrogate to NICMOS F110W) magnitudes. 

After flux scaling and astrometric registration (see SSH05), all suitable PSF subtracted images\footnote{Except those using HD 61005 possessing a bright light-scattering disk, and HD 25427 rejected due to instrumental image degradation.} showed similar: (a) scattered light excesses at r $<$ 1.0$^{\prime\prime}$, (b) oversubtractions at r$\gtrsim$1.2$^{\prime\prime}$. The latter is most likely result from chromatic effects in PSF subtraction.  With a J-H color index of +0.64, PDS 66 is quite red compared to $\sim$ +0.2 for G-F main sequence stars used as initial PSF templates. 

Better PSF subtractions were obtained using a high SNR, well color-matched, coronagraphic calibration PSF template derived from observations of the calibration star HD 82558 (spectral type K0, J = 6.08, J-H = +0.48) observed in HST GO program 10177 (PI: G. Schneider).  The two PDS 66 minus HD 82558 PSF subtracted images were median combined after rotating difference images to a common celestial frame and masking in each a small local artifact as well as the HST diffraction spikes. The scattered-light results discussed in this paper are derived exclusively from the two-orientation recombination of these higher fidelity PSF subtractions.

\subsection{ATCA Data}\label{atca}
The Australia Telescope Compact Array (ATCA) was used to observe the continuum emission towards PDS 66 in the 3 mm (94.1 GHz), 12 mm (18.9 GHz), 3 cm (8.6 GHz), and 6 cm (4.8 GHz) bands between 28 August and 2 September 2005.  The phase center was 13$^{h}$ 22$^{m}$ 8.53$^{s}$ in RA and -69$^{\circ}$ 38$^{\prime}$ 12.17$^{\prime\prime}$ in DEC (J2000), which is offset by 5.1$^{\prime\prime}$ in RA from PDS 66 to avoid artifacts created at phase center.  The ATCA interferometer consists of six 22-meter telescopes.  We used the H214 configuration, where 5 antennas are arranged with baselines between 82 and 247 m, and the sixth antenna is on a 4.5 km baseline.  The atmospheric phase stability was too poor to calibrate the longest baseline, and PDS 66 is too faint for self-calibration, and therefore data from the sixth antenna were discarded.  The system temperature for the single sideband receiver ranged from 380-500 K at 3mm, and 30-60 K at 12mm, 3cm, and 6cm.  The correlator was configured for 32 channels across 128 MHz of bandwidth in each band.  We used the MIRIAD package \citep{sault} to reduce and calibrate our data, and all of the calibration sources are from the 1997 ATCA Calibrator Source Catalogue by J. E. Reynolds\footnote{ftp://ftp.atnf.csiro.au/pub/atnfdocs/guides/at.cat}. 

\begin{figure}[ht]
\hspace{-0.2in}
\vspace{0.1in}
\includegraphics[scale=0.35,angle=-90]{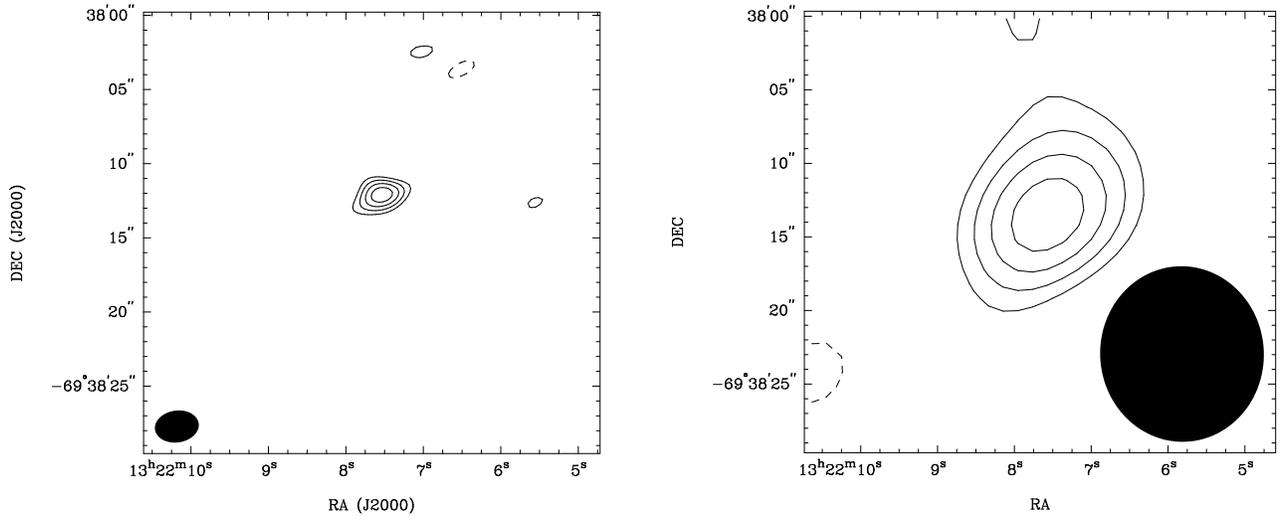}
\hspace{0.6in}
\vspace{0.1in}
\includegraphics[scale=0.35,angle=-90]{f1b.eps}
\caption{3mm (left) and 12mm (right) emission from PDS 66.  The fwhm of the synthesized beam is shown at the bottom of each plot in solid black.  The contours are drawn as n$\sigma$, where n=-6,-5,-4,-3,3,4,5,6 and $\sigma$=3.2 mJy/beam for 3mm and 0.06 mJy/beam for 12mm.  The negative contours are dashed and the positive contours are solid.}\label{plotone}
\end{figure}

The bandpass, flux, and gain calibrators for the observations in each of the four bands are listed in Table \ref{obs_table} with other key observational parameters.  Observations in the 3 mm band were obtained on four separate days. The map presented in Figure \ref{plotone} was produced from one day of the observations which had the best atmospheric phase stability. We measured a flux density of 2.2 Jy at both frequency windows (93.1 and 95 GHz) for 1057-797 on 2 September 2005 using Uranus as the flux calibrator. We assumed 1057-797 had a constant 3 mm flux over the four days.  It was used as a secondary flux calibrator for the other three days.  The flux scale in the 12 mm band was set by averaging the flux of 1352-63 over two runs, which lowered our uncertainties.

\begin{table}[ht]
\parbox{0.9\textwidth}{\caption{ATCA Observing Parameters\label{obs_table}}}
\centering
\begin{tabular}{l | c c c c}
\hline\hline
 & 3 mm & 12 mm & 3 cm & 6 cm\\
  \hline
Primary Flux Calibrator & Uranus\tablenotemark{a} & 1934-638 & 1934-638 & 1934-638\\
\hspace{0.1in} Assumed Flux Density (Jy)  & 4.5 & 0.99 & 2.8 & 5.8 \\
\hline
Phase Calibrator & 1057-797 & 1352-63 & 1251-713 & 1251-713 \\
\hspace{0.1in} Derived Flux Density (Jy)\tablenotemark{b}& 2.2$\pm$0.3 & 0.98$\pm$0.03 & 0.94$\pm$0.03 & 1.02$\pm$0.03\\
\hline
Bandpass Calibrator\tablenotemark{c} & 1921-293 & 1253-055 & 1253-055 & 1253-055 \\
\hline
PDS 66 & & & &\\
\hspace{0.1in} Derived Flux Density (mJy) & 22.0$\pm$3.3 & 0.41$\pm$0.06& $<$0.20\tablenotemark{d} & $<$0.25\tablenotemark{d} \\
\hspace{0.1in} Internal Uncertainty (mJy) & 2.4 & 0.06 & &\\
\hspace{0.1in} Calibration Uncertainty (mJy) & 2.3 & 0.01 & &\\
\hline
UV Range (k$\lambda$)& 21-69 & 3-16 & 1.5-7 & 0.5-4 \\
\hline
FWHM Primary Beam ($^{\prime}$) & 0.5 & 2.5 & 5.5 & 9.9 \\
\hline
FWHM Synthesized Beam ($^{\prime\prime}$) & 3.1$\times$2.2 & 10.2$\times$9.2 & 23.8$\times$20.2 & 43.5$\times$34.9 \\
\hline
\hline
\end{tabular}
\tablenotetext{a}{The flux of Uranus was derived based on an assumed brightness temperature and a uniform disk.}
\tablenotetext{b}{The uncertainties were estimated by measuring the flux at different elevations during the track.}
\tablenotetext{c}{The bandpass is the frequency-dependent gains}
\tablenotetext{d}{3$\sigma$ upper limits}
\end{table}

\begin{figure}[ht]
\begin{center}
\includegraphics[scale=0.6,angle=90]{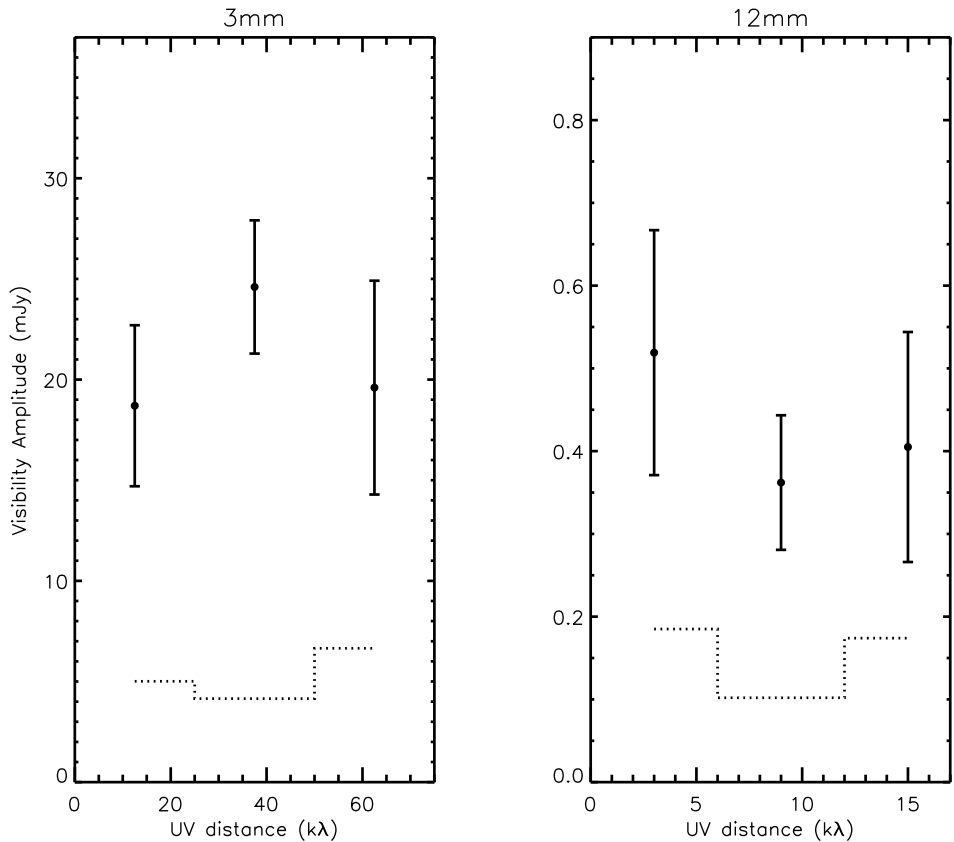}
\caption{Visibility amplitude as a function of projected baseline length for PDS 66 observations in the 3mm (left) and 12mm (right) bands.  The dotted line is the expected amplitude for zero signal, and is calculated in MIRIAD based on the observed variance within each bin.  The data points in each bin are the azimuthally averaged visibilities.  Since the profile is consistent with being flat, the disk of PDS 66 is not spatially resolved in these observations.}\label{plottwo}
\end{center}
\end{figure}

Table \ref{obs_table} lists the derived flux densities or the upper limits for PDS 66 in each of the four bands, along with the uncertainties (internal, calibrated, and total).  The signal-to-noise ratios for 3 and 12 mm are 6.7 and 6.8, respectively.  There was no detection at 3 or 6 cm, and so we list their 3-$\sigma$ upper limits.  The fluxes were measured in the UV-domain using UVFIT with a point source model. Images were made using INVERT with a robust weighting of 2 to optimize sensitivity, and the CLEAN algorithm was used to deconvolve the images.  The restored maps in Figure 1 are shown for 3mm (left) and 12mm (right), with the fwhm of the beam shown in the bottom left corner.  The 3mm data provide the maximum angular resolution of 2.2$^{\prime\prime}$ or 190 AU at the distance of PDS 66 (86 pc).  The visibility amplitudes as a function of projected baseline length is shown in Figure \ref{plottwo}.

\clearpage

\subsection{Spectral Energy Distribution}\label{lit}
\begin{figure}[ht]
\begin{center}
\includegraphics[scale=0.6,angle=90]{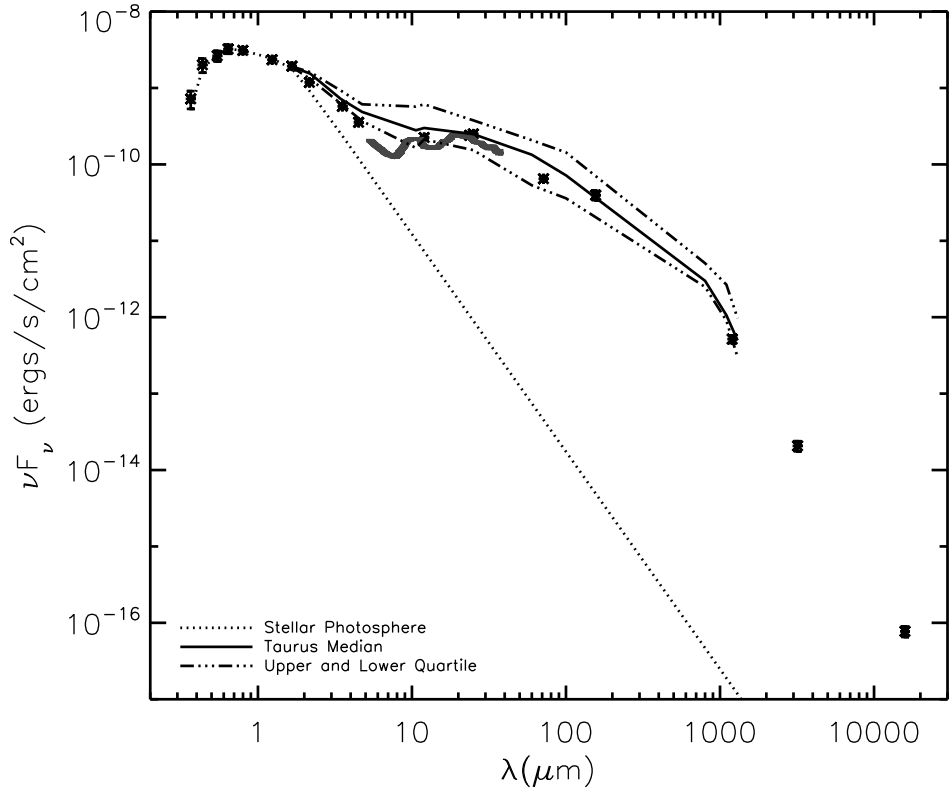}
\caption{Dereddened spectral energy distribution of PDS 66 composed of data from the literature, the low-resolution IRS spectrum from \cite{bouw08}, and new millimeter points from ATCA described in \S \ref{atca}.  A K1IV star with T$_{eff}$=5035 K (colors from \cite{KH95}) and normalized to I-band is shown as the dotted line.  The Taurus mean SED normalized to H-band is shown for reference as the solid black line, with the upper and lower quartiles in dash-dot lines \citep{dal99}. }\label{plotthree}
\end{center}
\end{figure}

The SED for PDS 66 shown in Figure \ref{plotthree} was derived from data obtained with several instruments, taken at different times.  The U,B,V,R$_{c}$,I$_{c}$ photometry are taken from the mean of \cite{bat}, since they show that PDS 66 varies by 0.1 mag over 5.75 days at these wavelengths.  We include this photometric variability as a source of uncertainty in the optical photometry, and find that the fluxes agree with the B and V photometry from the Tycho-2 catalog \citep{hog}.  We use 2MASS J,H,K photometry \citep{skrut06}, and IRAS 12 and 25 $\micron$ photometry.  We did not include the IRAS 60 $\micron$ data point because the field was contaminated by cirrus.  The Spitzer data, including IRAC, MIPS, and a low-resolution IRS spectrum, were obtained and reduced by the Formation and Evolution of Planetary Systems (FEPS) team \citep[][and references therein]{meyer06, bouw08}.  The 1.2 mm point comes from the Swedish-ESO Submillimeter Telescope \citep{carp08} and the 3 and 12 mm data are those described in \S \ref{atca}.  

We estimate the extinction towards PDS 66 by comparing the dereddened spectrum to a standard spectrum for a K1IV star from \cite{KH95} normalized to I-band.  We found that extinction corrections derived from color excesses in (B-V) overestimates the fluxes at $\lambda$ longer than I-band, and underestimates the fluxes at $\lambda$ shorter than I-band; while color excesses in (J-H) shows the opposite trend.  We find the best agreement for extinction as estimated from color excess with (R-I)$_{c}$ and (V-J), which are in good agreement with each other.  We adopt $A_{v}$=0.7$\pm$0.2 mag.  This is different than the value derived in \cite{MML02}, where they used the weighted mean of (B-V) and (V-J) to derive $A_{v}$=0.17$\pm$0.07.  The reason for the discrepancy might be due to the slightly elevated B-band fluxes produced by accretion onto the star, instead of arising from stellar emission alone.  We update the luminosity accordingly to 1.1 $L_{\sun}$, utilizing the bolometric correction from \cite{KH95} for a K1V star.  This higher luminosity decreases the age of PDS 66, assuming the same temperature (5035 K) as \cite{PM}.  Using the stellar evolutionary models of \cite{baraffe98}, we estimate an age of 11 Myr.  These same models were used to estimate the age of the moving group (13 Myr) to which PDS 66 belongs \citep{PM}, this mean group age is more robust than the age derived from the position of PDS 66 on the HR-diagram alone, the results are consistent within the uncertainties.   We corrected the flux densities for extinction using the extinction law from \cite{mathis90} translated into the Johnson-Cousins color system.  Color corrections were applied to the IRAC (IRAC Data Handbook, v. 3.0) and MIPS (MIPS Data Handbook, v. 3.2) photometry.  Color corrections were not applied to the IRAS data because the correction factors were close to unity (within 3\%).  The total uncertainties represent the internal and absolute calibration errors, and extinction corrections where appropriate.  The literature data are listed in Table \ref{lit_table}, and the dereddened SED is shown in Figure \ref{plotthree}.

The SED of PDS 66 is compared to the median Taurus SED \citep{dal99} in Figure \ref{plotthree}, where we have normalized to the H-band.  One should recall that:  (1) the SpT of the Taurus sources used to construct the median are K5-M2, whereas the SpT of PDS 66 is K1, and (2) PDS 66 is likely more face-on than the mean expected inclination of the Taurus median (which is probably closer to 60$^{\circ}$ assuming a random distribution of inclination angles).  With these caveats, we note the flux deficit relative to the Taurus median in the 4-20 $\micron$ region.  This could be the result of disk-clearing close to the star, or shadowing of parts of the disk further out in radii by the puffed-up rim.  A detailed analysis accounting for the inclination and SpT is described in \S \ref{radius}.  Another feature which stands out from the Taurus median is the flux deficit near 70 $\micron$.  Since PDS 66 is more face-on than the expected inclination of the median Taurus disk, a higher FIR luminosity is expected \citep[see Figure 5A in][]{ddn01}.  On the other hand, the SpT is earlier for PDS 66, resulting in a larger scale height of the PDS 66 rim, so more shadowing of neighboring parts of the disk would lower the FIR emission \citep[see Figure 5C in ][]{ddn01}.  However, the difference between the PDS 66 and the Taurus median SED may be a real physical signature of a less flared disk with larger grains.  Discussion of these possibilities is presented in \S \ref{unusual}.   

\begin{table}[ht]
\centering
\begin{tabular}{c c c}
\hline
$\lambda(\micron)$ & $F_{\nu}$ (Jy) & $\sigma_{total}$ (Jy) \\
\hline\hline
  0.37 &  0.031 &  0.003 \\ 
  0.44 &  0.122 &  0.006 \\ 
  0.54 &  0.263 &  0.007 \\ 
  0.64 &  0.392 &  0.011 \\ 
  0.80 &  0.546 &  0.010 \\ 
  1.24 &  0.791 &  0.023 \\ 
  1.66 &  0.909 &  0.018 \\ 
  2.16 &  0.789 &  0.013 \\ 
  3.54 &  0.657 &  0.014 \\ 
  4.51 &  0.521 &  0.012 \\ 
  7.74 &  0.471 &  0.010 \\ 
 12.0 &  0.882 &  0.053 \\ 
 23.7 &  1.874 &  0.120 \\ 
  25.0 &  2.071 &  0.145 \\ 
 71.5 &        1.54 &       0.12\\
 156.0 &        2.06 &       0.29\\
  1200 &       0.207 &      0.02\\
       \hline
\end{tabular}
\caption{Reddened fluxes for PDS 66 from the literature, as described in the text.}\label{lit_table}
\end{table}

\clearpage 

\section{Analysis and Results From the Scattered-Light Disk}\label{scatanalysis}
The HST/NICMOS 1.1 $\micron$ scattered-light image of the PDS 66 circumstellar disk is shown in Figure \ref{{plotfour}} (right image).  A faint, point-like, object apparently superimposed upon the disk at a celestial PA of 359$^{\circ}$, 1.2$^{\prime\prime}$ from PDS 66 (F110W mag of $\sim$17.2; $\delta$-F110W $\sim$+9.0) appears in the NICMOS image.  VLT/NACO J, H, K images obtained in May 2003 also reveal the object, but not the disk (M. Janson, priv. comm. May 2006).  The VLT images were used in combination with the NICMOS observations and establish through differential proper motions measures that this is a background star and not a faint companion associated with PDS 66.

The morphology of the 1.1 $\micron$ PSF-subtracted two-orientation combined scattered-light coronagraphic image  of the PDS 66 disk suggests an inclined disk with a PA=169$\pm$6$^{\circ}$ for the disk major axis.  The disk is robustly detected on the major axis to r$\sim$1.5$^{\prime\prime}$, with a low surface brightness scattering component extending to r$<$2$^{\prime\prime}$ ($\sim$170 AU).  The disk appears mirror symmetric (at low spatial frequencies) about the minor axis.  The eastern hemisphere of the disk (to the left in Figure \ref{{plotfour}}) is significantly brighter  (at equal radii) then the opposing side of the disk at all diametrically opposed azimuth angles, suggesting preferential forward scattering toward the morphological minor axis at PA$\sim$79$^{\circ}$.  The azimuthal brightness asymmetry with respect to the apparent major axis of the disk is indicative of an inclined disk. Assuming the apparent eccentricity is due to projection, the inclination is estimated to be 32$\pm$5$^{\circ}$ from face-on based upon the apparent major and minor axis lengths (mean axial ratio=1:0.848) determined from five equally spaced isophotal fits from r=12 to 22 pixel (0.91$<$r$<$1.67$^{\prime\prime}$).

We measure the 1.1 $\micron$ scattered light flux density of the PDS 66 circumstellar disk from an effective inner working angle of 0.41$^{\prime\prime}$ (one resolution element beyond the 0.3$^{\prime\prime}$ coronagraphic mask radius) to 3.5$^{\prime\prime}$ (beyond which no significant disk-scattered starlight contributes to the total brightness) as 2.7$\pm$0.4 mJy.  This excludes a small area circumscribing the coronagraphic obscuration (shown as black in Fig \ref{{plotfour}} and rejected as unreliable due to imperfections in PSF subtraction very close to the star).  Thus, the total disk flux density may be somewhat greater.  We adopt an F110W absolute instrumental sensitivity of 1.26$\times$10$^{-6}$ Jy per ADU/sec/pixel and a Vega system photometric zero point of 1775 Jy for a spectrally flat source.  By scaling the brightness of the HD 2558 template  to the PDS 66 observations, we find the F110W magnitude of PDS 66 to be $\sim$8.29 (= 860 mJy).  Thus, the fraction of 1.1 $\micron$ starlight scattered by the disk at r$>$0.41$^{\prime\prime}$ is approximately 0.32\%. 

While the PDS 66 disk appears bilaterally symmetric (implying a disk inclined to the line of sight), we measured, and fit, the disk's radial surface brightness profile medianed over all azimuth angles (Figure \ref{{plotfour}}).  We find that within 0.8$<$r$<$1.5$^{\prime\prime}$, the disk's azimuthally medianed surface brightness, SB(r) declines as 370.2$\times$r$^{-4.53} \mu$Jy/arcsec$^{2}$ (goodness of fit R=0.993). 

\begin{figure}[ht]
\begin{center}
\includegraphics[scale=0.4]{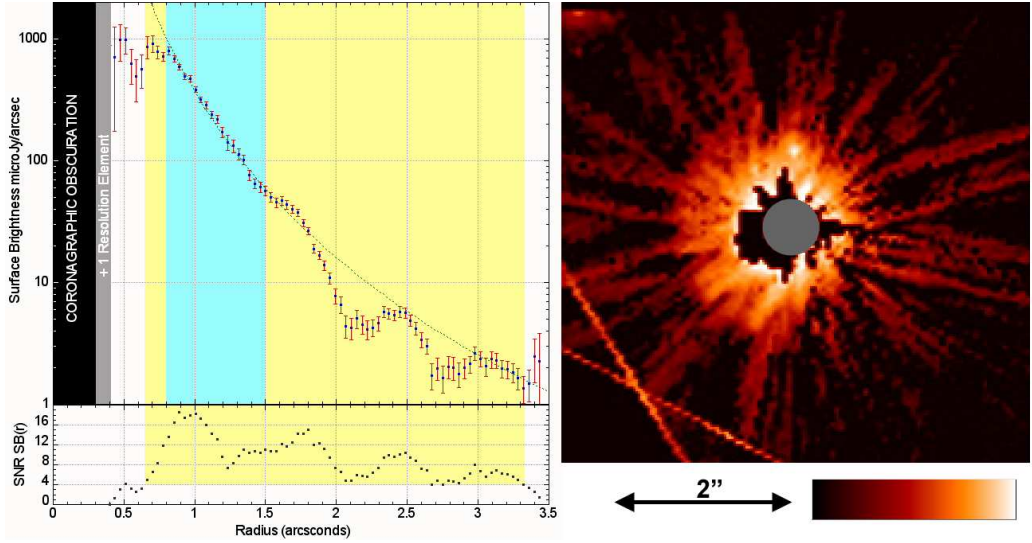}
\caption{NICMOS F110W 2-roll combined PSF-subtracted imaging observations of the PDS 66 circumstellar disk.  Left: Azimuthally medianed radial surface brightness profile (excluding unreliable data close to the star, shown masked as black in plot).  Error bars are standard deviations (1$\sigma$) about the median surface brightness (SB) at corresponding radii. Dotted line is extrapolation of a power law fitted in the region in blue (0.8$^{\prime\prime}<$r$<$1.5$^{\prime\prime}$) extrapolated to r=3.5$^{\prime\prime}$. Plotted below is the SNR corresponding to radial flux density measurements in each one pixel wide annular zone.  Right: 7.5$\times$7.5$^{\prime\prime}$ field (north up, east left) centered in the NICMOS coronagraph; log10 stretch from +0.001 to +10 ADU s$^{-1}$ pixel$^{-1}$ (1 ADU s$^{-1}$ pixel$^{-1}$ = 218 $\mu$Jy arcsec$^{-2}$).  Filled gray circle (r=0.41$^{\prime\prime}$) is one resolution element beyond image plane obscuration.}\label{{plotfour}}
\end{center}
\end{figure}
\clearpage

\subsection{Grain Properties Inferred from Millimeter Wavelengths}\label{graingrowth}
The growth of grains may be inferred from the slope of the millimeter spectrum, where the emission goes as $\nu^{\alpha}$.  Assuming the Rayleigh-Jeans limit, and that the opacity is a simple power-law going as $\nu^{\beta}$, it follows that $\alpha=2+ \beta$ if the emission is optically thin.  Smaller grains (sizes $\ll\lambda$, e.g. ISM-like grains) have $\beta\sim$ 2, while larger grains (sizes $\gg\lambda$) have $\beta\sim$ 0 and the emission follows the shape of a blackbody.  Potentially complicating this simple SED analysis,  cTTS occasionally show signs of ionized stellar winds, which can dominate the centimeter emission with spectral slopes ranging from $\nu^{-0.1}$ \citep{mez67} for optically thin to $\nu^{0.6}$ \citep{pan75, WB75, ol75} for optically thick emission.  The stellar wind can contribute significantly to emission in the millimeter, so corrections for a wind component must be applied so that only thermal dust emission is considered in deriving limits on the spectral slope.  \cite{rod06} fit the centimeter emission for young stars in Taurus, and found a range of slopes from $\nu^{-0.1}$ to $\nu^{0.4}$.  We use our $3\sigma$ upper limits from the centimeter observations with the adopted range of slopes from \cite{rod06} to estimate possible contributions from stellar winds.  

Figure \ref{{plotseven}} shows the mm-wave spectrum, with a linear least-squares fit to the data points (dotted black line), and estimates of the stellar wind (dashed and dotted lines) going through the upper limits to the centimeter data.  Although we do not find evidence for a strong stellar wind, we point out that the values extrapolated from the 3$\sigma$ upper limits could contribute significantly at 12mm (between 43-65\%).  However, since we did not detect such emission at 3 or 6 cm, we assume that stellar wind is negligible.  

We measure a slope of 2.4$\pm$0.1 for the millimeter emission, which corresponds to a $\beta$ of 0.4$\pm$0.1.  This derivation of $\beta$ assumes purely optically thin emission, yet there can be a non-negligible contribution from optically thick emission which would cause $\beta$ to appear lower.  The correction to $\beta$ to account for some optically thick emission, defined using $\Delta$, is described in detail in \cite{beck90} (see their equation 20).  We follow their same definitions for the temperature, surface density, and opacity profiles; and their same normalization of 0.1 cm$^{2}$/g at 10$^{12}$ Hz (0.3 mm) for the total (dust+gas) opacity.

$\Delta$ was derived assuming that the density and temperature decrease with radius as $r^{-p}$ and $r^{-q}$, respectively, and so it depends on the density and temperature profile indices, $p$ and $q$.  $\Delta$ is weakly dependent to R$_{disk}$, M$_{disk}$, and $i$ as they are captured in the ln$\bar{\tau}$ factor (Equation 16 in \cite{beck90}). Since we only have crude estimates of these quantities, we can only approximate the contribution to $\beta$ from optically thick emission.  We generously estimate $\Delta$ assuming $\kappa_{1.2 mm}$\footnote{$\kappa_{1.2 mm}$ maximizes the correction, since $\kappa_{3mm}$ or $\kappa_{12mm}$ lead to a lower value for $\Delta$.}, $p$=1.5, $q$=0.75, $R_{disk}$=170 AU\footnote{Higher $R_{disk}$ values will lead to lower values of $\Delta$.} (see \S \ref{scatanalysis}), $M_{disk}$=0.005 $M_{\sun}$ \citep{carp05}, and $i=32^{\circ}$ (see \S \ref{scatanalysis}).  The value for $p$ is at the high end of a reasonable range according to \cite{AW07}; and $q$ is from the optically thin, geometrically flat disk discussed in \cite{adams87}, although typical values for $q$ are lower than this \citep[e.g.][]{beck90}.   We find $\Delta\sim$0.2 and $\beta_{corr}\lesssim$ 0.5.  

This low value of $\beta$ may indicate particle sizes near the mm-regime according to the models of \cite{MN93}, where they show that $\beta\lesssim$1 for a grain size distribution of $n(a)\propto$a$^{-3.5}$ and a maximum grain size of 1mm.  More recent models by \cite{dal01} also suggest that such low values of $\beta$ are found if the maximum grain radii are in the millimeter to centimeter regime.  Since the millimeter emission traces the bulk of the disk mass, it is possible to infer the disk mass from the flux at these wavelengths.  \cite{carp05} derive a dust mass for the disk of PDS 66 of 5$\times$10$^{-5}$ M$_{\sun}$, and our 3 and 12 mm fluxes give very similar masses even with our new estimate of $\beta$.  Note that the derived total dust mass may be somewhat higher from the low surface brightness diffuse scattering seen beyond 75 AU in scattered light.

\begin{figure}
\vspace{-2in}
\begin{center}
\includegraphics[scale=0.5,angle=90]{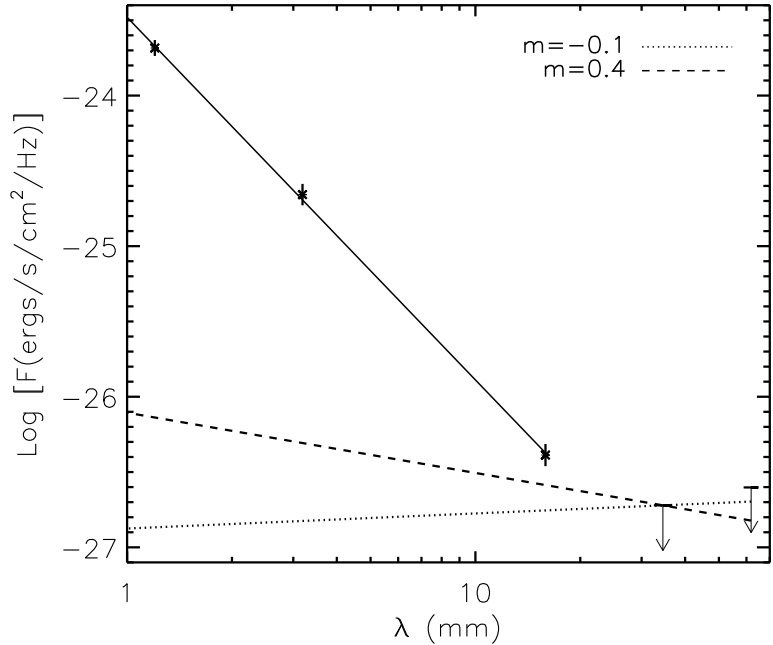}
\caption{Long-wavelength spectrum showing the fit to the observed data points (dotted line).  We measure a slope of -2.4$\pm$0.1, which corresponds to a $\beta$ of 0.4$\pm$0.1.  The solid and dashed lines represent typical slopes for the stellar wind \citep{rod06} that has a frequency dependence $\nu^{m}$, and are constrained by our 3 and 6 cm upper limits (arrows).  We assume that the effects of stellar wind are probably insignificant.  $\beta$ has been corrected for optically thick emission, using values for the temperature and density power-laws such that the correction is maximized.  We find $\beta_{corr}\leq$ 0.5.}\label{{plotseven}}
\end{center}
\end{figure}

\clearpage

\section{Analysis and Results From SED Models}\label{results}
Detailed modeling of the SED combined with resolved imagery at one or more wavelengths can lead to a better understanding of the physical structure of the circumstellar environment \citep[e.g.][]{schn03}.  In theory, one can learn about the composition and size of the grains, the disk geometry, and mass surface density as a function of radius.  However, we are limited in the parameters we can derive since the disk was not resolved in the millimeter.  For example, the outer disk radius, surface density profile, and disk mass are degenerate parameters, since we cannot distinguish between a steep mass distribution concentrated tightly around the star and a shallow mass distribution extending further out.  However, we can still constrain specific physical parameters in regions of the SED which are not strongly dependent on degenerate model parameters. We focus on estimating the inner disk radius, inferring the degree of flaring, and constraining of the surface density and disk mass.  We estimate the inner dust truncation radius by fitting a single temperature blackbody to the near infrared (NIR) emission (2-5 $\micron$), described in \S \ref{radius}.   The fixed model parameters throughout this analysis are the stellar mass (1.3 M$_{\sun}$), effective stellar temperature (5035 K), and the distance to the system (86 pc).  We choose between models in \S \ref{radius} using the $\chi^{2}$ statistic.  Penultimately, we use our millimeter observations to place constraints on the surface density and disk mass in \S \ref{flatdisk} using a flat disk model.

\subsection{Inner Disk Radius}\label{radius}
A large fraction of the flux in the NIR can be dominated by emission from the inner rim (significantly inside the inner working radius of the NICMOS coronagraph which is 25 AU at 86 pc), a region where the local disk opacity changes dramatically.  The change in opacity is caused by the boundary between temperatures that are above and below dust sublimation, and the lack of dust inside of this boundary results in a drop to zero opacity \citep{muz03}.  The inner edge of the dust disk is in full view of the star, and therefore, the temperature is higher than it otherwise would have been if the dust disk were not truncated.  The gas in the disk near the point of dust sublimation is also hotter, and so the disk scale height increases there because the pressure is greater.  But, the disk might also appear to have a region cleared of dust if the grains have grown more in the inner regions, since larger grains have lower opacities. Faster growth is expected closer in because the particles have higher Keplerian velocities and the surface density in the disk is higher, thus higher collision rates are expected there. A planet forming in the disk would also decrease the opacity in its vicinity because it would sweep up the surrounding material, creating a gap.

It is interesting to test if any of the above scenarios are consistent with the observations of PDS 66, which is possible by estimating the temperature (and thus location) of the inner rim.  We calculate its emission as $F_{\nu}$=B$_{\nu}\times$area \citep[following][]{ddn01}, where the area is a function of the inclination (greater surface area of the inner rim wall is seen for more edge-on disks) and vertical scale height of the rim.  The U, B, V, R, I, J, H, and K stellar colors are estimated from \cite{KH95} assuming a K1V star; and we extrapolate the fluxes beyond K-band assuming $F_{\nu}\propto\nu^{2}$.  The central luminosity, inclination, and vertical scale height are fixed at  1.1 $L_{\sun}$ (see \S \ref{lit}), 32$^{\circ}$ (see \S \ref{scatanalysis}), and 0.1 $R_{rim}$, respectively.  Since a constraint on the vertical scale height is not available, it is the dominant source of uncertainty in $R_{rim}$.  Varying the vertical scale height from 0.1$R_{rim}$ to 0.25$R_{rim}$ \citep{ddn01} leads to a change in $R_{rim}$ by a factor of 1.5.

We fit the temperature and solid angle of the inner rim for the 1-5 $\micron$ photometry only, assuming the rim emission is a Planck function.  We do not expect the 8 $\micron$ point to fit well with a single temperature blackbody at the inner rim, as it is likely contaminated by silicate emission and cooler dust temperatures.  The left plot of Figure \ref{{ploteight}} shows the fit for the rim temperature and radius which give the minimum $\chi^{2}_{\nu}$:  $T_{rim}$=1277 K and $R_{rim}$=0.07 AU.  The right plot of Figure \ref{{ploteight}} shows the reduced $\chi^{2}$ as a function of the rim temperature, where the temperature grid explored exceeds the range shown on this plot (900-1400 K).  We find that the rim temperature is within 1190-1370 K for a 3$\sigma$ confidence level, and the probability of these data being drawn from a model with this temperature is 24.5\%.    This result indicates that the dust temperature and inner rim radius are consistent with being located at the dust sublimation radius.

\begin{figure}[ht]
\begin{center}
\includegraphics[scale=0.8,angle=90]{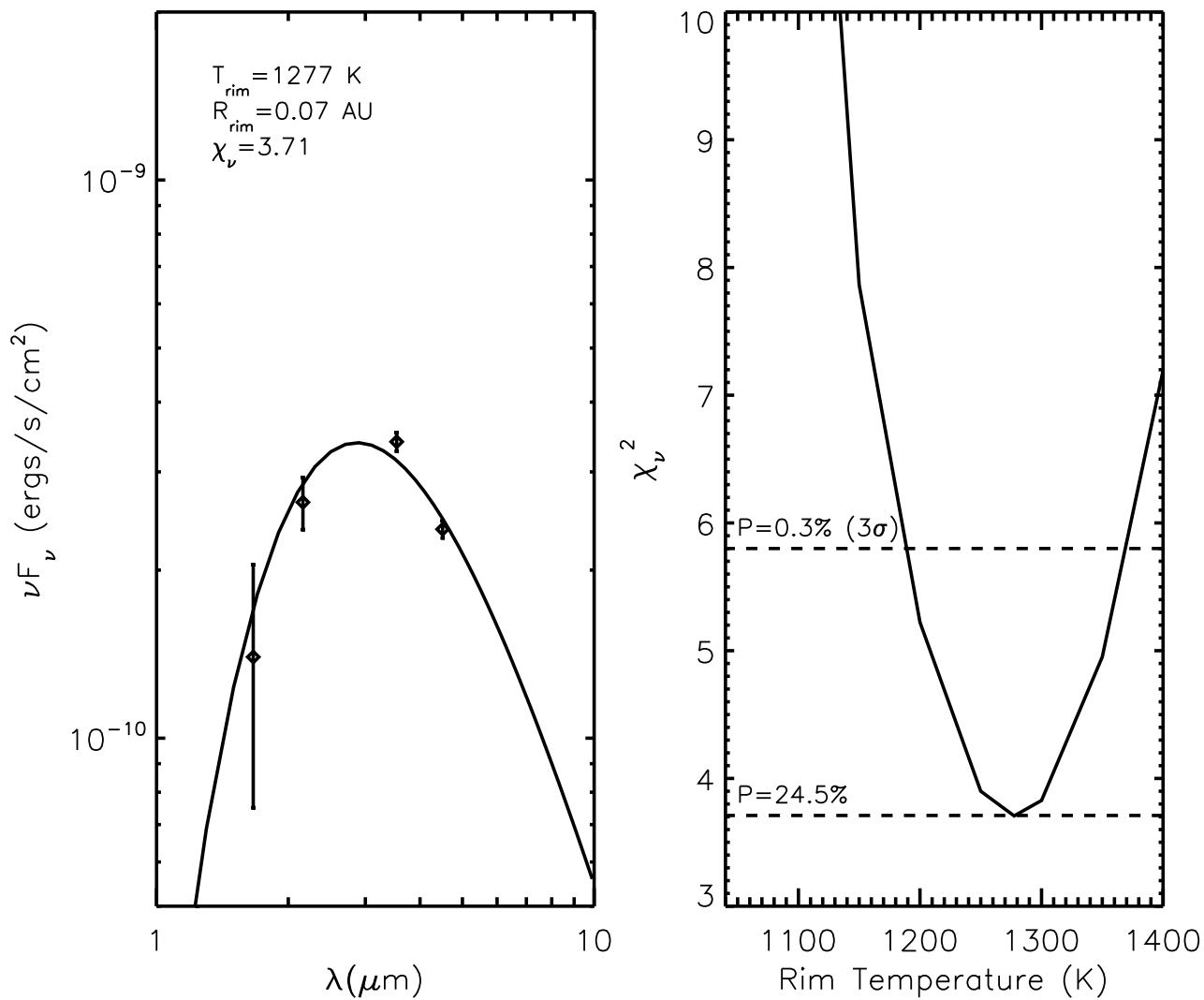}
\caption{The left plot shows the best-fitting rim temperature (1277 K) and radius (0.07 AU) for a single temperature blackbody (solid line) model.  The black diamonds are the data points for the excess (disk) emission: J, H, K, 3.6, and 4.5 $\micron$, and their error bars include the internal, calibration, and extinction uncertainties.  Note that we did not include the 8 $\micron$ data point in the fit.  Plotted on the right is the reduced $\chi^{2}$ as a function of rim temperature, showing that the temperatures within 3$\sigma$ that are fit by the model.}\label{{ploteight}}
\end{center}
\end{figure}

\clearpage
\subsection{Flat Disk Model}\label{flatdisk}
In the millimeter regime, it is a reasonable approximation to model the disk as geometrically flat because, at these wavelengths, the emission is optically thin and we can observe the midplane of the disk where the particle density is the highest.  We are interested in learning about the possibility of planets forming in this disk, and particularly if the surface density and disk mass are sufficient.   The following equations can be used to estimate the surface density in the planet forming region, where the normalization is at 5 AU.  

We assume a radial temperature distribution with power-law index, $q$.  We take the temperature from the best-fitting single-temperature blackbody to the NIR emission described in \S\ref{radius}: $T_{rim}$=1277 K at 0.07 AU, and extrapolate to larger radii assuming $q$=0.5.  This value of $q$ was assumed by fitting the SED between 11-70 $\micron$ following Equation 10 in \cite{beck90}.  The total surface density distribution is given by 
\begin{equation}
\Sigma(r) = \Sigma_{5} \left(\frac{r}{5 AU}\right)^{-p}
\end{equation}
We assume $p=1$ for the bulk of this discussion, but consider values for $p$ ranging between 0 and 1.  The assumption of $p=1$ and the range in $p$ is motivated by millimeter observations that suggest the mean value of $p$ observed in disks is close to unity and most disks fall within the range 0$\leq$p$\leq$1 and the \citep[e.g][]{AW07, kit02}.  The surface density normalization, $\Sigma_{5}$, is evaluated at 5 AU; and for reference, the minimum mass solar nebula \citep[MMSN,][]{hayashi81} is $\Sigma_{5}$=150 g/cm$^{2}$ for gas and dust.  Although the value for the MMSN assumes $p=3/2$, which  is large compared to observed values of $p$, the normalization determined by \cite{hayashi81} is substantially larger than is constrained by our observations. 

The quantities described above can be used to estimate specific parameters related to the emission coming from a disk.  In particular, the flux for a geometrically flat disk is given by
\begin{equation}
F_{\nu}=\frac{\mu}{d^{2}}\int_{R_{in}}^{R_{out}}B_{\nu}(1-e^{-\tau_{\nu}/\mu})2\pi rdr
\end{equation}
$B_{\nu}$ is the Planck function, $\tau_{\nu}$ is the optical depth, $\mu$=cos\textit{i}, and $d$ is the distance to the source.  Since we are considering emission at wavelengths beyond 1 mm, we can assume the Rayleigh-Jeans limit ($T<$30 K), and that the disk is optically thin:  (1 - e$^{-\tau_{\nu}/\mu})\approx$ $\tau_{\nu}/\mu$=$\kappa_{\nu}\Sigma(R)/\mu$.  The opacity, $\kappa_{\nu}$, is the same equation as described in \cite{beck90}.  We use an inner disk radius of 0.07 AU from the fit to the NIR.  With the given assumptions, we can solve for the surface density at 5 AU:
\begin{equation}\label{{ploteleven}_eqn}
\Sigma_{5} = \frac{F_{\nu}c^{2}d^{2}(2-p-q)}{\nu^{2}\kappa_{\nu}4\pi kT_{0}r_{0}^{q}r_{5}^{p}(R_{out}^{2-p-q} - R_{in}^{2-p-q})}
\end{equation}
Figure \ref{{ploteleven}} shows $\Sigma_{5}$ as a function of outer disk radius, ranging from 1 to 200 AU, calculated with Equation \ref{{ploteleven}_eqn}, and using the 3mm flux.  We consider a range for the power-law on the surface density between 0 and 1.  The vertical line labeled `$\tau_{vis}\geq10^{4}$ yrs' is the size of a disk which has been growing due to viscous evolution for 10$^{4}$ yrs \citep{holl05}.  If material is moving inwards, and eventually accreting onto the star, the disk spreads out on a viscous timescale ($\tau_{vis}=r^{2}/\alpha$) to conserve angular momentum.  The region to the right of the line represents disks that have higher viscous timescales, and therefore larger outer disk radii.  We choose a viscous timescale that is much shorter than the age of the star because there are other processes that may be interfering with disk spreading, making the viscous timescale longer.  Thus, we are setting a conservative lower limit (10 AU) to the outer disk radius, but we realize that the disk must be larger than this based on both the age of the star and the HST/NICMOS observations (\S \ref{scatanalysis}).  The line labeled `mm-beam' is the maximum beam size from the 3mm observations (130 AU), and we assume the disk size must be smaller since the emission from PDS 66 is unresolved (\S \ref{atca}).  It is clear in Figure \ref{{ploteleven}} that as the outer disk radius marches down, more mass must be packed into a smaller and smaller disk, and so the surface density at 5 AU increases rapidly.  If we consider the case of $p=1$ (dash-dot line) we require that 3 $<\Sigma_{5}<$ 11 g/cm$^{2}$ based on the radius constraints.  If the disk follows a shallower density profile, shown with the dashed and dotted lines in Figure \ref{{ploteleven}}, the spread between upper and lower limits on $\Sigma_{5}$ increases.

The disk mass is tied to the surface density via the equation $M=\int_{disk}2\pi r\Sigma(R)dr$, and so a higher surface density leads to a higher disk mass.  If we integrate over the disk, the total disk mass is given by 
\begin{equation}
M_{d} = \frac{2\pi\Sigma_{5}r_{5}^{p}(R_{out}^{2-p}-R_{in}^{2-p})}{2-p}
\end{equation}
The total (dust+gas) disk mass derived by assuming $p=1$ is 0.0004 $<M_{d}<$ 0.001 M$_{\sun}$.  For comparison, the total disk mass derived in \cite{carp05} from their 1.2 mm observation is 0.005 $M_{\sun}$.  This is approximately 17 $M_{\earth}$ in dust mass, and 5.3 $M_{jupiter}$ in total mass (assuming a gas-to-dust ratio of 100).

Since it is more realistic that the power-law on the temperature and density profiles vary with disk radius, we describe how our results change if these indices are different.  One extreme may be that the disk is optically thick and geometrically thin ($q=0.75$) for all radii, making the derived disk mass higher by a factor of 2-5.  If instead, we keep the same temperature profile ($q=0.5$) but consider flatter density profiles (by changing the $\Sigma_{r}$ profile from $p=1$ to $p=0$), then the limits on the disk mass will increase by a factor of 1.25-2 for the range in radii we explore.

\begin{figure}
\begin{center}
\includegraphics[scale=0.6,angle=90]{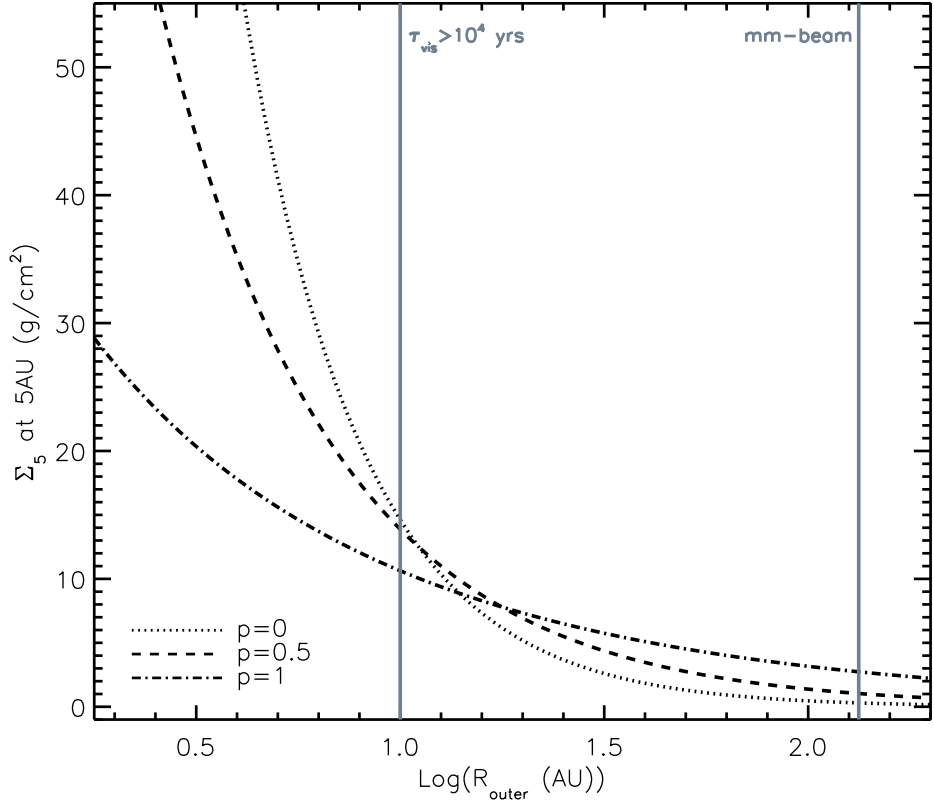}
\caption{Total (dust + gas) surface density at 5 AU as a function of outer disk radius, calculated according to Eqn \ref{{ploteleven}_eqn}, using our 3mm flux.  Shown are the results for three values of the power law on the surface density profile:  $p=0$, $p=0.5$, and $p=1$.  The vertical lines are the maximum disk size assumed from the mm observations and the lower limit on the outer disk radius is shown as the vertical line labeled `$\tau_{vis}>10^{4}$yrs', based on the size of the disk estimated from viscous evolution.  Assuming the $p=1$ model is correct, which is reasonable for a T Tauri star, we estimate the surface density at 5 AU and use it to constrain the total disk mass to 3$\leq\Sigma_{5}\leq$11 g/cm$^{2}$ and 0.0004 $<M_{d}< $0.001 M$_{\sun}$, respectively.}\label{{ploteleven}}
\end{center}
\end{figure}

\clearpage

\section{Discussion}\label{discuss}
\subsection{How Unusual is PDS 66?}\label{unusual}
Despite its age, PDS 66 ($\sim$13 Myr) is analogous to the well-studied cTTS in Taurus (1-3 Myr).  Compared to the sizable sample of stars in Taurus studied by \cite{furlan06}, the spectral indices computed between 13 and 25 $\micron$ vs. 6 and 13 $\micron$ for PDS 66 ($n_{6-13}=-0.13$, $n_{13-25}=0.36$) are typical of the flared, Class II sources (see their Figure 11).  Figure \ref{plotthree} also illustrates the similarity between the SED of PDS 66 and the median SED of T Tauri stars in Taurus.  Most sources older than 10 Myr do not display primordial signatures, yet PDS 66 appears similar to younger cTTS.  There may be hints that PDS 66 is slightly more evolved than the mean disk in Taurus, but proper accounting for biases between the two SEDs must be taken before drawing conclusions.  We used a simple model fit to the NIR emission discussed in \S \ref{radius} to show that the inner rim is similar in temperature and location to values for cTTS derived in other works \citep[e.g.][]{muz03, meyer97}.  The outer disk radius assumed from the HST/NICMOS scattered light image ($\sim$170 AU) is consistent with the mean outer disk radius derived from mm-observations \citep{AW07}.  Although, mm- and NIR scattered light-derived outer disk radii tend to differ.  The outer disk radius of IM Lup, for example, is 400 AU derived from NIR scattered light \citep{pin08} and only 105 AU from resolved mm-observations \citep{lom07}.  This discrepancy is probably due to different sensitivities to mass surface density and emission that traces different grain populations. 

We now compare PDS 66 to other sources of comparable in age.  TW Hydra, a 10 Myr old cTTS, appears to have an inner hole that is 4 AU from the star, well beyond the dust sublimation radius, inside of which, there is only optically thin material \citep[$\tau_{\nu}\lesssim0.05$, ][]{calvet}.  This inner hole might be created by a planet sweeping up material from the disk.  \cite{set08} claim that radial velocity variations in TW Hydra could be caused by a planet, but subsequent observations and analysis suggest that the variations are instead caused by cool star spots \citep{huel08}.  The disk mass of TW Hydra ($\sim$0.009 $M_{\sun}$, \cite{krist00}) is within a factor of 2 of PDS 66 (0.005 M$_{\sun}$), however the mass accretion rate of TW Hydra ($5\times10^{-10} M_{\sun}/yr$, \cite{muz00}) is lower by an order of magnitude.  The mass accretion rate may be a means of discriminating between cTTS and transitional disks, as is discussed in \cite{naj07}, where transitional disks have lower accretion rates for a given mass.  TW Hydra falls in the transition disk territory, having a lower mass accretion rate for its mass.  Interestingly, PDS 66 falls on the border in between cTTS and transitional disks for single stars, hinting that it may be evolving towards the transition disk population.  Perhaps the differences between PDS 66 and TW Hydra are related to the initial conditions, such as available circumstellar material.  That could explain why TW Hydra appears to be more evolved at a similar age, at least in terms of the inner disk properties and accretion rate. 

\cite{AW07} find that the mean surface density at 5 AU for their sample disks in Taurus (ages $\lesssim$7Myr) is $\sim$14 g cm$^{-2}$, which is comparable to the values we estimate here with $p=1$ and $q=0.5$.  The total disk mass constrained by our millimeter observations is on the low end of the distribution for disks in Taurus, which may be a sign of disk dissipation over time.  The value we derive for $\beta$ is smaller than the mean for Taurus, where $\beta\sim1$, and therefore PDS 66 may have a larger grain population, on average.  TW Hydra and CQ Tau are examples of two other long-lived disks that also appear to have grains that are several millimeters in size \citep{wilner05, testi}.  

\cite{bouw08} give a detailed analysis of the low-resolution IRS spectrum for PDS 66, and they find evidence for grain sizes up to 5 $\micron$.  Their observations are probing the surface layers of the disk, where smaller grains spend most of their time.  Evidence of even larger grains are probed by our millimeter observations, but these grains are found primarily in the midplane of the disk.  Although our observations and those of  \cite{bouw08} are probing different regions, it is clear that the two disk layers are connected.  Figure 7 of \cite{lom07} shows that there is a correlation between the strength/shape of the 10 $\micron$ feature and the slope, $\alpha$, of the submillimeter emission.  Observations of PDS 66 are consistent with this correlation, with $\alpha=-2.4$, and the normalized $F_{11.3}/F_{9.8}\sim0.97$ and peak $F_{10\micron}\sim1.5$ \citep{bouw08}.  Although the statistics are still limited, the interpretation of this trend is that there is a depletion of sub-$\micron$ sized particles because they stick together and eventually grow to larger particles.  During this grain growth process, the 10 $\micron$ feature flattens out and the slope of the millimeter spectrum becomes shallower, due to the changing opacity.  We expect that larger grains will settle to the midplane, causing the disk to appear less flared.  This is exactly what is seen by \cite{bouw08} (their Figure 6), where the slope of the FIR emission is correlated with the strength of the 10$\micron$ feature (or, as just mentioned, the slope in the submillimeter).  Since the slope of the FIR emission decreases as the disk becomes less flared, it is clear that increased grain growth, traced by either the 10$\micron$ feature or the millimeter slope, leads to more settling, and a less flared disk.  Therefore, our observation of a steeper mm-slope is consistent with the flattened 10 $\micron$ feature found by \cite{bouw08}, this is most likely accompanied by a less flared disk geometry.

\subsection{Planet-Forming Potential of PDS 66}\label{planform}
It is not clear if giant planets form within all disks, since current statistics are still limited.  At least 80\% of the youngest pre-main sequence stars are thought to have circumstellar disks \citep[][]{hill98, haisch01, hern07}, whereas only 10.5\% of stars have gas giant planets with orbital periods $<$2000 days \citep{cumming08}.  One prerequisite to form giant planets by core accretion is sufficient disk mass in the region where they form, which takes precedence over the shape of the surface density profile \citep[e.g.][]{raymond05}.  On the theoretical side, core accretion models of planet formation require $<$5 Myr to form Jupiter with a core mass ranging between 5-16 $M_{\earth}$ \citep[e.g][]{hub05}.  If this timescale is right, then long-lived disks such as PDS 66 have had enough time to form a giant planet.  Higher mass stars have a greater chance of harboring planets at larger radii \citep{johnson07}, perhaps simply because more material is available; and supporting observations suggest that disk mass increases with stellar mass \citep{nat00}.  Could the disk of PDS 66 have enough mass to form planets?  

\cite{pasc07} observed NeII emission towards PDS 66 and that the star is actively accreting, providing ample evidence that there is still gas in the disk.  We turn to our mm data to address how much material may be available to form a planet.  As \cite{AW07} point out, the surface density constrained by observations is merely a snapshot in time, while the formation of planets is more likely dependent on the evolution of the disk mass (or surface density) in the region where planets might form.  If the estimated range in the surface density at 5 AU in \S \ref{flatdisk} are correct, we can compare that to the minimum requirements for planet formation from current models \citep[e.g][]{boss98,boss05,poll96,inaba03,dur05}.  Figure 17 of \cite{AW07} shows the cumulative mass requirements to form giant planets from current models.  We also find that our upper limit on the total disk mass that is too low to form giant planets by about an order of magnitude from the minimum mass requirement.  We cannot rule out, however, that a planet may have already formed, though there is no evidence of gaps having developed in the SED associated with a forming giant planet.  A limitation of our analysis is that our unresolved millimeter observations are integrated over the disk, and so cannot constrain the mass in potential planet forming regions of the disk.  Future observations offering significant improvement in spatial resolution in the posited planet-forming region, such as with ALMA, will improve constraints on the surface density in the regions where planets would form.

\section{Summary}\label{summ}
We have carried out a detailed analysis of the $\sim$13 Myr old cTTS, PDS 66, using data from 0.1 $\micron$ out to 12 mm.  We conclude the following:

1.  Analysis of the HST/NICMOS scattered light imagery reveals a near-IR surface brightness radial distribution going as $r^{-4.53}$, detected out to 2$^{\prime\prime}$ or 170 AU.  The inclination (32$\pm5^{\circ}$) was derived from the asymmetry in the disk brightness, assuming a mean axial ratio$=$1:0.848.  These parameters can be used to remove ambiguity in future modeling efforts that could be partially constrained by the surface brightness profile.

2.  The power-law on the opacity spectrum is $\beta\leq$ 0.5, which was derived from the slope of the millimeter spectrum and reflects a maximized correction for optically thick emission.  The low value of $\beta$ suggests grain growth from the assumed initial ISM sizes, and theoretical models suggest that $\beta\lesssim$1 implies millimeter sized grains. 

3.  Using a single-temperature blackbody (assuming the inclination from our scattered light imaging), we fit the near-IR portion of the SED by varying the inner rim radius and temperature.  We find that the best-fit parameters are $R_{rim}$=0.07 AU and $T_{rim}$=1280 K, consistent with inner dust disks of other cTTS being truncated at the dust sublimation temperature.

4.  Using a geometrically flat, optically thin disk model, we estimate the total surface density at 5 AU to  3 $<\Sigma_{5}<$ 11 g/cm$^{2}$, which leads to constraints on the total disk mass of 0.0004$< M_{d}<0.001 M_{\sun}$.  This is in accordance with a previous total (dust+gas) disk mass estimate \citep{carp05} of 0.005 $M_{\sun}$ (corresponding to 17 $M_{\earth}$ in dust and 5.3 $M_{Jupiter}$ in gas, assuming a dust-to-gas ratio of 100).  These mass constraints indicate that this disk is unlikely to form a Jupiter mass planet now, although we found that variations in the surface density profile or opacity can lead to disk masses that are higher by as much as a factor of 5.  This result does not rule out that the possibility that a planet may already have formed, though there is no evidence to support this.
 
\section*{Acknowledgments}
We are grateful to the anonymous referee for providing useful suggestions that greatly improved this paper.  We thank the FEPS team for their valuable contributions and encouragement.  SRC thanks E. Mamajek for his useful suggestions and making us aware of his unpublished work, which both improved the quality of this paper.  We also thank C.P. Dullemond for providing CGPLUS.  SRC wishes to thank St\'{e}phane Herbert-Fort for many useful discussions.  The Australia Telescope Compact Array is part of the Australia Telescope which is funded by the Commonwealth of Australia for operation as a National Facility managed by CSIRO.  This work is based, in part, on observations with the NASA/ESA Hubble Space Telescope, which is operated by AURA, under NASA contract NAS5-26555.  The HST observations are associated with GO programs 10527 and 10177 with support provided by NASA grants through STScI.  MRM, IP, and SRC gratefully acknowledge support from NASA Astrobiology Institute through LAPLACE.  FEPS was supported through NASA contracts 1224768, 1224634, and 1224566, managed by JPL.

{\it Facilities:} \facility{ATCA}, \facility{HST (NICMOS)}

\end{document}